\documentclass[usenatbib,useAMS]{mnras}

\usepackage{graphicx}
\usepackage{bm,amssymb,amsmath}
\usepackage{times}
\usepackage{multicol}
\usepackage{float}
\usepackage{color}
\usepackage{cancel}
\usepackage[normalem]{ulem} 
\graphicspath{{./fig/}{./png/}{./ps/}}

\newcommand{\vect}[1]{{{\mbox{\boldmath $#1$}}}}		

  \newcommand{\vv}{{_{\rm h}}}  				
  \newcommand{\hh}{{_{\rm m}}}  				

  \newcommand{\Mm}{\,{\rm Mm}}

  \newcommand{\K}{\,{\rm K}}

  \newcommand{\G}{\,{\rm G}}
  
  \newcommand{\mT}{\,{\rm mT}}

  \newcommand{\GO}{{G}}

  \newcommand{\BO}{B_{0z}}

  \newcommand{\bc}{{b_{00}}}

  \newcommand{\fO}{{f_{0}}^2}

  \newcommand{\GFME}{{Paper~I}} 
  \newcommand{\GFE}{{Paper~II}} 
  \newcommand\leftidx[3]{%
  {\!\vphantom{#2}}\,#1\hspace{-0.05cm}#2#3%
}
  \definecolor{burntorange}{RGB}{255,97,0}
  \definecolor{violet}{RGB}{105,20,200}
  \definecolor{royalblue}{RGB}{0,137,255}
  \definecolor{mygreen}{RGB}{0,120,0}

\definecolor{midblue}{rgb}{0.0,0.4,0.7}
\definecolor{purple}{rgb}{0.4,0.2,0.5}
\newcommand{\rev}[1]{\textcolor{black}{#1}}

  \title[Magnetic network construction]{
    {Modell}ing 3D magnetic networks in a realistic solar atmosphere}
  \author[Gent et al.]{
    Frederick~A.~Gent$^{1,3,4}$\thanks{
    E-mails: frederick.gent@aalto.fi
    }, 
    Ben~Snow$^{2}$, Viktor~Fedun$^5$,
    Robertus~Erd\'{e}lyi$^{4,6}$\\
    $^1$Academy of Finland ReSoLVE Centre of Excellence, 
        Department of Computer Science, Aalto University FI-02150\\
    $^2$College of Engineering, Mathematics and Physical Sciences, University of Exeter,
    EX4 4QF, UK\\
    {$^3$School of Mathematics, Statistics and Physics, Newcastle University, NE1 7RU, UK}\\
    $^4${Solar Physics and Space Plasma Research Centre (}SP$^2$RC), School of Mathematics and Statistics, University of Sheffield, S3 7RH, UK \\
    $^5$Plasma Dynamics Group, Department of Automatic Control and Systems Engineering, University of Sheffield, S1 3JD, UK \\
    {$^6$Department of Astronomy, E\"otv\"os Lor\'and University, P\'azm\'any P\'eter s\'et\'any 1/A, H-1117 Budapest, Hungary}
    }
  
\begin{document}

  \maketitle
  \begin{abstract}
    The magnetic network extending from the photosphere (solar radius $\simeq R_\odot$) to lower corona ($R_\odot+10\Mm$) play{s} an important role in the heating mechanism{s of the solar atmosphere}.
    Here we {further develop} the models {with realistic open magnetic flux tubes} of 
    {the authors} in order to model more complicated configurations.
    Closed magnetic loops, and combinations of closed and open magnetic flux tubes are modelled.
    These are embedded within a stratified atmosphere, {derived from observationally motivated semi-empirical and data-driven models} subject to solar gravity and {{capable of spanning from the photosphere up into} the chromosphere and lower corona.}
    Constructing a magnetic field comprising self-similar magnetic flux tubes, an analytic solution for the 
    kinetic pressure and plasma density is derived.
    Combining flux tubes of opposite {polarity} it is possible to create a steady background 
    magnetic field configuration modelling {a} solar atmosphere {exhibiting realistic stratification}.
    The result can be applied to {SOHO/MDI and SDO/HMI and other}  magnetograms from the solar surface, 
    upon which photospheric motions can be simulated to explore the mechanism of energy 
    transport.
    We demonstrate th{is powerful and versatile} method with an application to Helioseismic and Magnetic Imager data. 
  \end{abstract}


  \section{Introduction}\label{Intro}

  Since \rev{the} discovery that the solar corona \rev{is} significantly hotter than the photosphere, following the {1932} solar eclipse \citep{CM35} and subsequent confirmation \citep{Redman42}, {how so} has {posed} a major challenge.
  {Across} the solar atmosphere temperatures vary by orders of magnitude{. T}ypical photospheric temperatures are about $6500\K$ (solar radius $R_\odot\simeq696\Mm$), and above $10^6\K$ in the corona (out to about $2R_\odot$) \citep[][and references therein]{Priest87, EP14, Asch05, Ebook08}.
  The solar surface and atmosphere are extremely dynamic.
  Frequent and powerful events such as coronal mass ejections release high energy, localised heating within the atmosphere, and yet the corona everywhere is hot.
  Jets, flares, prominences, and flux emergence, among others, carry mass and energy from the surface into the atmosphere. 
  However, it remains unclear how energy is {dissipated} through the chromosphere
  and subsequently to the coronal plasma \citep{Zirker93,Asch05,Klimchuk06,DeP11,vBATDeL11,PCS18,Zetal18}.
  Persistent and ubiquitous small{-}scale processes would appear to be candidates for this effect.
  Some advocate small{-}scale reconnections of magnetic field lines
  {\citep{GN02,GN05,BBP13}}.
  An alternative view may be that solar magnetic field lines, in the form of magnetic \emph{flux tubes}, act as guides for magnetohydronamic (MHD) waves that may carry the missing energy to heat the atmosphere to observed temperatures.
  These occur at scales, which are increasingly available to observational comparison \citep[]{JMMKAB07,MVJKRME12,WSSRLFE12}.
  This article is motivated by the latter, but may nevertheless be useful more generally.

  {While m}odels of magnetic field configuration dealing with coronal heating often set the flux-tube footpoints {at the photosphere, some discount the effects of the chromosphere and} \rev{the} Transition Region (TR), a relatively narrow layer between chromosphere and corona where \rev{there is a jump in} plasma \rev{density and} temperature.
  In the corona the magnetic field is commonly modelled as force-free {\citep[for example]{SdR03,SSAdR04}}, assuming the plasma pressure to be negligible, but in the {low} chromosphere and the photosphere kinetic forces cannot reasonably be ignored, with the ratio of thermal to magnetic pressure plasma-$\beta\gg1$.
  The dynamic interface region (IR) {includes the chromosphere and TR, connecting} the photosphere and lower corona {\citep{LMSAL14}}.
  Typical mass and energy density in the IR are orders of magnitude larger than in the corona as a whole \citep{MTW75,VAL81,FATH06,FBH07,FCHHT09}, so it is reasonable to expect IR dynamics to be critical for the coronal heating mechanism.

  The extreme nine orders of magnitude gradient in plasma density (six in pressure, three in temperature) over $2.5\Mm$ from the upper photosphere to the lower corona presents a significant challenge in modelling magnetic fields in the chromosphere \citep{Def07}.
  Typical magnetic flux-tube footpoint strength of about $100\mT~ (1000\G)$ are observed emerging from the photosphere
  \citep[][and references therein, the latter Ch.8.7, Ch.5, respectively]{Zwaan78,Priest87,EP14,Asch05, Ebook08}.
  An isolated magnetic flux tube must, therefore, expand exponentially in radius as it rises to balance the plasma pressure.
  {Although the solar atmosphere is highly dynamic and turbulent{,} many features, such as loops, spots
  and pores, can be observed to remain static for hours, days or even weeks}
  \citep{MTKWSS77,LW77,MSRM83} and this has been used to investigate the transport mechanisms along the field line{s} with a series of numerical simulations {\citep{SFE08,FES09,SZFET09,FSE11,VFHE12,KC12,MFE15,ME15}}.
  These {numerical studies} were restricted to single flux tubes and did not breach the TR, so flux tube interaction and the effect on the corona cannot feasibly be explored.
  \citet{KCF08,KC12} constructed a 2D magnetic field with multiple flux tubes, each identical to its neighbour, but excluding the TR.
  \citet{HvBKS05,HvB08} constructed a 2D magnetic field which does extend into the low corona.
  \citet[][hereafter \GFE]{GFE14} generalised the background configuration to 3D, multiple, non-identical flux tubes, extending into the lower corona.
  This was {successfully} applied to a 3D model of a flux-tube pair by \citet{SFGVE18}, who showed that chromospheric shocks at the intersections between the tubes are capable of driving supersonic jets.

  However, all of these models apply only to open magnetic flux tubes of the same polarity.
  Their major omission is flux loops with footpoints of opposite polarity, which are common features of solar magnetic networks.
  \citet{VAU79} considered an analytic construction of a single 3D magnetic flux loop as a static background, but for a thermodynamic model, not MHD.
  The primary contribution of the current work will be to add loops to the multiple flux tube network described in \GFE.
  An advantage of this result shall be that any arbitrary magnetogram of the photosphere, e.g. from the Helioseismic and Magnetic Imager for SDO (HMI) \citep{HMI07}, can be constructed by matching the vertical field for each pixel to the model and constructing analytically a realistic 3D {magnetic} network extending into the corona.
  Using the corresponding velocity field from the same observational image or similar, forward modelling can then be applied to explore the energy transport mechanism.
  The analytical model is outlined explicitly in Section\,\ref{sect:loop},
  and differences with \GFE.
  In Section\,\ref{sect:apps} some applications for the model are described
  and some discussion of its uses and limitations.
  
  \rev{In general $\beta\ll1$ in the corona, so modelling perturbations about the steady background magnetic network without kinetic effects is reasonable. At the photosphere and in the lower chromosphere, however, $\beta>1$, except inside the low-$\beta$ sunspots and flux-tube footpoints. This is why it is important to model the steady background with kinetic and magnetic forces in equilibrium. In this framework we can examine the MHD processes localised around the strong magnetic structures, while on the time and length scales of interest the kinetically dominated ambient atmosphere supporting the magnetic structures also remains steady. Solving only the  perturbation fields can reduce the numerical challenges. Even with more complicated magnetic networks, various analytic photospheric flows can be applied to investigate how energy propagates and is dissipated through the magnetic network, to help identify the most relevant physical processes.}

  \section{Magnetic flux loop}\label{sect:loop}
  \subsection{Ambient magnetic field outside the flux tubes}\label{subsect:strat}
%
  In \citet[][hereafter \GFME]{GFME13a} we constructed analytically a 3D model
  of a single vertical magnetic flux tube
  embedded in a realistic solar atmosphere
  at magnetohydrostatic (MHS) equilibrium.
  This was extended to multiple magnetic flux tubes in \GFE.

  The background atmosphere employed was
  derived from the combined modelling profiles of
  \citet[][Table~12, VALIIIC]{VAL81}
  and \citet[][Table~3]{MTW75} for the chromosphere and lower solar
  corona, respectively (see Fig\,1\,\GFME).
  {We are only considering MHD, so require profiles only for the gas density and pressure or temperature to solve the steady state momentum equation.
  The atmospheric models selected are sufficient for the qualitative results of interest to us, but other
  profiles accounting for additional or specific physics would also work, providing they depend only on 
  solar radius.
  If \rev{additional} physics \rev{were included}, such as ionisation, radiative transfer, self-gravity, etc., or if the hydrostatic equilibrium depended on horizontal forces, an alternative solution \rev{would be} required.}

  Observations \citep[Ch.3.5 in][]{M93book,TS03} indicate \rev{that} the atmosphere outside the flux tubes includes a non-zero magnetic field of order $1-10\mT$ in the corona.
  It is important to model this ambient field, so that
  realistic ratios can be obtained between
  the thermal and magnetic pressures, i.e. plasma-$\beta<1$
  outside the flux tube.
  {\GFME\, and \GFE\, implemented explicit external fields to provide ambient magnetic pressure.}
  In this article we model magnetic flux loops
  by combining vertical flux tubes of opposite polarity.
  For a flux tube of opposite polarity an ordered ambient field will negate the effective field in the flux tube.
  Therefore, a constant vertical ambient field is not suitable for use with flux loops.
  A realistic solution still requires a low plasma-$\beta$ in the
  corona.
  {Further refining} the model the ambient magnetic pressure felt by each individual flux tube is {now} induced by the superposition of its neighbouring flux tubes.
  Plasma-$\beta<1$ {above the photosphere} will be obtained due to the expansion of strong flux tubes and loops near the local network.
  Therefore, we drop the ambient field denoted by $\bc$ in Equation\,22 of \GFE.

\subsection{The MHD equations} \label{subsect:MHD}
  A full {outline} of the governing ideal MHD equations, which we would use to {describe} the environment in the solar atmosphere is provided in \citet[][Section\,2.2]{GFE14}.
  Our approach,
  following that of \citet{SFE08}, is to derive the system of
  equations governing the perturbed MHD variables by splitting 
  the variables $\rho$ (plasma density), $e$ (energy density)
  and $\vect{B}$ (magnetic field) into their background 
  and perturbed components
  \begin{equation}
    \label{eq:tilde}
    \rho = \rho_b + {\tilde{\rho}},
    \quad\quad    e = e_b + {\tilde{e}},
    \quad\quad    \vect{B} = \vect{B}_b + \tilde{\vect{B}},
  \end{equation}   
  where tilde denotes the perturbed portion and it is assumed $\rho_b,\,e_b$
  and $\vect{B}_b$ do not vary with time.
  When the time-independent momentum equation describing the background equilibrium is deducted, the modified form of the momentum equation governing the perturbed system is given by
  \begin{eqnarray}
    \label{eq:momf}
    \frac{\upartial \left[\left(\rho_b+\tilde\rho\right) u_i\right]}{\upartial t}
    +\frac{\upartial
    }{\upartial x_j}
    \left[
    \left(\rho_b+\tilde\rho\right) u_iu_j
    -\frac{\tilde{B_i}\tilde{B_j}}{\mu_0}
    \right]
    &&\\\nonumber
    +\frac{\upartial}{\upartial x_i}
     \tilde p_T 
    -\frac{\upartial}{\upartial x_j}
    \left[
    \frac{\tilde{B_i}B_{bj}+B_{bi}\tilde{B_j}}{\mu_0}
    \right]
    +F_{{\rm bal}_i}
    & =& \tilde\rho g_i,
  \end{eqnarray}
  and the consequent energy equation is given by
  \begin{eqnarray}
    \label{eq:energyf}
    \frac{\upartial \tilde{e}}{\upartial t}
    +\frac{\upartial}{\upartial x_j}
    \left[
    (e_b+\tilde{e}) u_j 
    -\frac{\tilde{B_i}\tilde{B_j}}{\mu_0}u_i
    \right] &&\\\nonumber
    +\frac{\upartial}{\upartial x_j}
    \left[
    \tilde p_T u_j 
    -\frac{\tilde{B_i}B_{bj}+B_{bi}\tilde{B_j}}{\mu_0}u_i
    \right]
    &&\\\nonumber
    +p_{bT}\frac{\upartial u_j}{\upartial x_j} 
    - \frac{B_{bj}B_{bi}}{\mu_0} \frac{\upartial u_i}{\upartial x_j}
    +F_{{\rm bal}_i}u_i
     & =&
    \tilde\rho g_i u_i,
  \end{eqnarray}
  in which $\vect{u}$ and $\vect{g}$ are the velocity and
  gravitational acceleration.
  {$\vect{F}_{\rm bal}$ represent net background equilibrium forces.}
  The system is completed by the equations of continuity, induction and
  state, as detailed in \GFE.

  Given no vertical current $J_z$ a {stationary} state, where magnetic force balances exactly pressure and gravitational forces, has {an MHS} equilibrium solution providing the magnetic field satisfies
  \vspace{-0.4cm}\begin{equation}\label{eq:suff2}
    \upartial_y B_z
    \upartial_z B_x
    =
    \upartial_x B_z
    \upartial_z B_y,
  \end{equation}
  and, hence, $\vect{F}_{\rm bal}{=\vect{0}}$.
  A scalar solution for pressure can otherwise still be derived by inclusion of minimal horizontal balancing forces $\vect{F}_{\rm bal}$, yielding forced magnetohydrostatic equilibrium (FME).
  These {balancing forces} are small compared to the other forces and may be considered to be a statistical steady superposition of small-scale high-cadence turbulence in the chromosphere, where the magnetic field is not force-free.
  Our approach is to specify the background magnetic field. We then solve the time-independent momentum equation
  \begin{equation}\label{eq:delP}
  \vect\nabla p_b
  + \vect\nabla \frac{|{\vect{B}_b}|^2}{2\mu_0}
  -\left({\vect{B}}_b\cdot\vect\nabla\right)
             \frac{{\vect{B}}_b}{\mu_0} 
  -\rho_b g\vect{\hat{R}} + \vect{F}_{\rm bal}
  = 
  \vect{0},
  \end{equation}
  to find the {FME} $p_b$ and $\rho_b$, and identify the balancing forces $\vect{F}_{\rm bal}$.
  Gravity depends only on solar radius $\vect{{R}}$, 

  \subsection{A single magnetic flux tube}\label{subsect:01}
  In cylindrical coordinates, taking $\vect{\hat{z}}$ to be {along} $\vect{{R}}$, the magnetic potential of a self-similar axisymmetric magnetic flux tube is
  \begin{equation}
    \label{eq:potential}
    \leftidx{^m}A_{br} = \leftidx{^m}S\phi\, 
    \leftidx{^m}G
    \frac{\upartial \leftidx{^m}f}{\upartial r},\,    \leftidx{^m}A_{b\phi} = 0,\,    \leftidx{^m}A_{bz} = \leftidx{^m}S\phi\, 
    \leftidx{^m}G 
    \frac{\upartial \leftidx{^m}f}{\upartial z},
  \end{equation}
  or in Cartesian coordinates we have
  \begin{align}\label{eq:Axyz}
    \leftidx{^m}A_{bx}
    &=
    \leftidx{^m}S\arctan
    \left(
    \frac{y-\leftidx{^m} y\,}{x-\leftidx{^m} x\, } 
    \right)
    {\leftidx{^m}\GO \, }
    \frac{\upartial \leftidx{^m}f \,}{\upartial x}
    ,\nonumber
    \\
    \leftidx{^m}A_{by}
    &=
    \leftidx{^m}S\arctan
    \left(
    \frac{y-\leftidx{^m} y\,}{x-\leftidx{^m} x\, } 
    \right)
    {\leftidx{^m}\GO \, }
    \frac{\upartial \leftidx{^m}f \,}{\upartial y}
    ,
   \nonumber \\
    \leftidx{^m}A_{bz}
    &=
    \leftidx{^m}S\arctan
    \left(
    \frac{y-\leftidx{^m} y\,}{x-\leftidx{^m} x\, } 
    \right)
    {\leftidx{^m}\GO \, }
    \frac{\upartial \leftidx{^m}f \,}{\upartial z},
  \end{align}
  where $\leftidx{^m}{\vect{A}_b}$ denotes the potential for the $m^\textrm{th}$ flux tube, which has its axial vertical magnetic field $\leftidx{^m}S$ located at a footpoint $(\leftidx{^m}x,\leftidx{^m}y)$ on the photosphere.
  We {scale} $\leftidx{^m}f$ and $\leftidx{^m}G$ from \GFE\ to
  \begin{equation}
    \leftidx{^m}f=-\frac{\leftidx{^m}r^2{\BO}^2}{2}\, \textrm{  and  }\, 
    \leftidx{^m}G=\exp\left(\frac{\leftidx{^m}f}{\fO}\right), 
  \end{equation}
  with factor $\fO$ governing the radial scale of the flux tube and the radial distance $\leftidx{^m}r $ from the axis at 
  $(\leftidx{^m}x,\leftidx{^m}y)$ is 
  \begin{equation}
    \label{eq:ri}
    \leftidx{^m}r \, = \sqrt{(x-\leftidx{^m}x )^2+(y-\leftidx{^m}y )^2}.
  \end{equation}
  The reduction in the vertical field strength along the flux tube axis is specified by an appropriate monotonically decreasing function $\BO(z)$, such as a sum of exponentials as applied in \GFE\ or a polynomial form as applied by \citet{Gary2001} and
  employed in Section\,\ref{sect:arb}
  The sign of real parameter $\leftidx{^m}S$ determines the polarity of the flux tube.
  {The components of the magnetic field for the $m^{\textrm th}$ flux tube $\leftidx{^m}{\vect{B}_b}$ are then defined as in Equation\,22 of \GFE\ with $\bc=0$.
  Now, however, by} construction at $(\leftidx{^m}x,\leftidx{^m}y)$ $\leftidx{^m}G=1$ and $\leftidx{^m}r = \leftidx{^m}f =0$.
  We also impose $\BO(z=0) = 1$.
Hence, at the flux tube axis the photospheric magnetic field is 
  $\leftidx{^m}B_{bz} = \leftidx{^m}S$, which can be set directly or interpolated from HMI data or similar.

  Equation\,\eqref{eq:delP} can be decomposed into hydrostatic (HS) and MHS parts{, i.e.}
  \begin{eqnarray}\label{eq:delP1bh}
  \vect\nabla (p_b\vv +\leftidx{^m}{p}_b\hh 
                 )
  + \vect\nabla \frac{|\leftidx{^m}{\vect{B}_b}|^2}{2\mu_0}
  -\left(\leftidx{^m}{\vect{B}}_b\cdot\vect\nabla\right)
             \frac{\leftidx{^m}{\vect{B}}_b}{\mu_0} 
  &&
  \\
  \nonumber
  +\leftidx{^m}{\vect{F}}_{\rm bal} 
  -(\rho_b\vv +\leftidx{^m}{\rho}_b\hh
  ) g\vect{\hat{z}}
  &=& 
  \vect{0},
  \end{eqnarray}
  in which ${p}_b\vv$ and ${\rho}_b\vv$ denote HS plasma 
  pressure and density, and $\leftidx{^m}{p}_b\hh$ and $\leftidx{^m}{\rho}_b\hh$ denote MHS adjustments due to flux tube $\leftidx{^m}{\vect{B}}_b$.
  $\leftidx{^m}{\vect{F}}_{\rm bal}$ vanishes, with
  Equation\,\eqref{eq:suff2} satisfied for the single flux tube. 
  The HS equilibrium is constructed using the VAL IIIC \citep{VAL81} temperature and density profile{s} to calculate a  pressure profile, using the ideal gas law. That is then differentiated vertically to produce a stable density profile, assuming constant gravity. The advantage of this method is that it allows the pressure and density fields to be corrected, after the MHS corrections have been applied, to exclude negative values.
  What remains of Eq.\,\eqref{eq:delP1bh} is 
  \begin{equation}\label{eq:delP1b}
  \vect\nabla \leftidx{^m}{p}_b\hh 
  + \vect\nabla \frac{|\leftidx{^m}{\vect{B}_b}|^2}{2\mu_0}
  -\left(\leftidx{^m}{\vect{B}}_b\cdot\vect\nabla\right)            \frac{\leftidx{^m}{\vect{B}}_b}{\mu_0} 
  -\leftidx{^m}{\rho}_b\hh 
   g\vect{\hat{z}}
  = 
  \vect{0}{.}
  \end{equation}
  The solution to Eq.\,\eqref{eq:delP1b} follows \GFME\, and \GFE, in the absence of terms defining an ambient magnetic field $\bc$, to yield 
  \begin{eqnarray}\label{eq:px}
    \!\!\!\!\leftidx{^{m}}{p}_b\hh 
    \!\!&=&\!\!
      \frac{\leftidx{^m}S^2
           }{2\mu_0}\leftidx{^m}\GO^2 
      \left[
           \fO\BO\BO^{\prime\prime} + 
         2\leftidx{^m}f{\BO^\prime}{}^2
           -{\BO}^4 
       \right],\\
   \label{eq:rho1}
    \!\!\!\!\leftidx{^m}{\rho_b\hh}\!\!&=&\!\!
    \frac{\leftidx{^m}S^2\leftidx{^m}\GO^2}{\mu_0g}
    \left[
      \left(\frac{\fO}{2}+2\leftidx{^m}f\right)\BO^\prime\BO^{\prime\prime}
      +\right.\\\nonumber
      \!\!&&\!\!\left.\qquad\qquad\quad\frac{\BO\BO^{\prime\prime\prime}{\fO}}{2}
      -2
      {\BO}^3\BO^\prime
    \right]
    .
  \end{eqnarray}
  \vspace{-0.5cm}
\subsection{Including a second or more flux tubes of mixed polarity} \label{sect:amb12}
  \begin{figure*}
  \centering
  \includegraphics[trim=3.8cm 0.4cm 3.0cm 1.4cm, clip=true, width=0.95\linewidth]{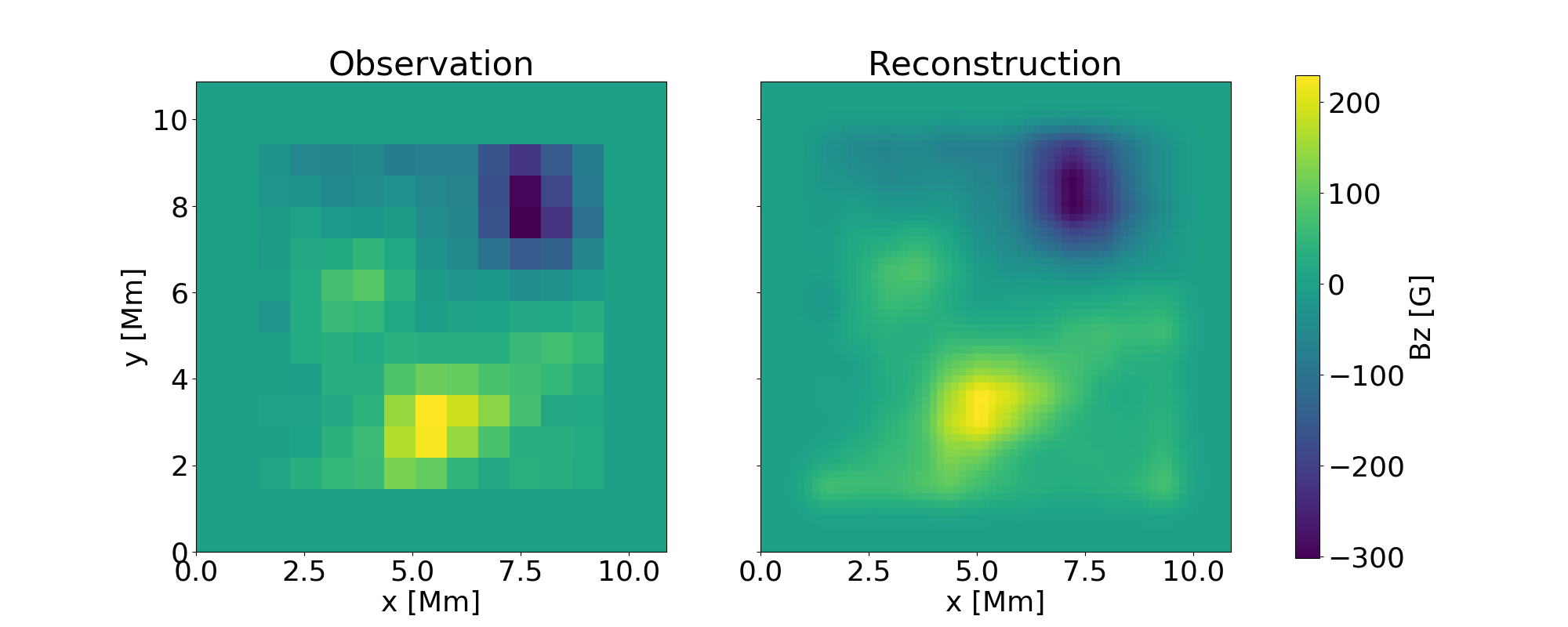}
  \caption{{Observed HMI magnetogram (left) model (right) of the photosphere}. {F}illed pixels highlight the resolution.
  \label{fig:hmimap}}
  \end{figure*}

  {Let us} now include a second flux tube, such that  $\leftidx{^n}{\vect{B}_b}$ denotes one with the same construction as $\leftidx{^m}{\vect{B}_b}$ apart from the arbitrary axial coordinates $(\leftidx{^n}x,\leftidx{^n}y)$ and parameter $\leftidx{^n}S$.
  Equation\,\eqref{eq:delP1b} becomes
  \begin{eqnarray}\label{eq:delP012}
  \vect\nabla (\leftidx{^m}{p}_b\hh 
                 + \leftidx{^{n}}{p}_b\hh
                 + \leftidx{^{mn}}{p}_b\hh)
  &&
  \\
  \nonumber
  -\left([\leftidx{^m}{\vect{B}}_b+\leftidx{^n}{\vect{B}_b}]\cdot\vect\nabla\right)
             \frac{\leftidx{^m}{\vect{B}}_b+\leftidx{^n}{\vect{B}_b}}{\mu_0} 
  + \vect\nabla \frac{|\leftidx{^m}{\vect{B}_b}+\leftidx{^n}{\vect{B}_b}|^2}{2\mu_0}
  &&
  \\
  \nonumber
  +\leftidx{^{mn}}{\vect{F}}_{\rm bal} 
  -(\leftidx{^m}{\rho}_b\hh 
  +\leftidx{^{n}}{\rho}_b\hh+\leftidx{^{mn}}{\rho}_b\hh) g\vect{\hat{z}}
  &=& 
  \vect{0},
  \end{eqnarray}
  where superscript $^{n}$ has equivalent meaning for the second flux tube as indicated for the first in Eq{uation}\,\eqref{eq:delP1bh}.
  The {additional} superscript $^{mn}$ {refers} to the interaction between the flux tube pair.
  Subtracting Equation\,\eqref{eq:delP1b}, and the equivalent for the second flux tube retains
  \begin{eqnarray}\label{eq:delP12}
  \vect\nabla 
        \leftidx{^{mn}}{p}_b\hh
  -\left(\leftidx{^m}{\vect{B}}_b\cdot\vect\nabla\right)
             \frac{\leftidx{^n}{\vect{B}_b}}{\mu_0} 
  -\left(\leftidx{^n}{\vect{B}_b}\cdot\vect\nabla\right)
             \frac{\leftidx{^m}{\vect{B}}_b}{\mu_0} 
  &&
  \\
  \nonumber
  + \vect\nabla \frac{\leftidx{^m}{\vect{B}_b}\cdot\leftidx{^n}{\vect{B}_b}}{2\mu_0}
  +\leftidx{^{mn}}{\vect{F}}_{\rm bal} 
  - 
  \leftidx{^{mn}}{\rho}_b\hh g\vect{\hat{z}}
  &=& 
  \vect{0}.
  \end{eqnarray}
  Equation\,\eqref{eq:suff2} is not satisfied, so $\leftidx{^{mn}}{\vect{F}}_{\rm bal}$ does not vanish. 
  \begin{eqnarray}
        \frac{\upartial }{\upartial x}
        {\leftidx{^{mn}}{p}_b\hh}
        &=&
     \frac{2\leftidx{^n}f ^2}{\fO}
      {\BO^\prime}^2
      \leftidx{^m}S
      \leftidx{^n}S
      {\BO}^2 \leftidx{^m}\GO 
      \leftidx{^n}\GO 
    \frac{x-\leftidx{^n}x}{\mu_0} 
    \\
    \nonumber&+&
    \frac{2\leftidx{^m}f^2}{\fO}
      {\BO^\prime}^2
      \leftidx{^m}S
      \leftidx{^n}S
      {\BO}^2 \leftidx{^m}\GO\leftidx{^n}\GO 
    \frac{x-\leftidx{^m}x}{\mu_0}
    \\
    \nonumber&+&
        \frac{\upartial }{\upartial x}
      \left(
      \frac{
      \leftidx{^m}S
      \leftidx{^n}S
      \fO
      }{2\mu_0}
      \leftidx{^m}\GO 
      \leftidx{^n}\GO 
      \left[
      {\BO^\prime}^2
      +\BO
      \BO^{\prime\prime}
      \right]
      \right){,}
  \end{eqnarray}
  {in which the fir}st two lines cannot integrate with respect to $x$, while a similar residual expression is obtained from integrating the $y$-component of Eq{uation}\,\eqref{eq:delP12}.
  However, a scalar solution for the pressure and density is possible, if this contribution to the magnetic tension force is balanced by 
  \begin{eqnarray}\label{eq:F12}
    \leftidx{^{mn}}{\vect{F}}_{\rm bal} &=&-
     \frac{2}{\fO}
      {\BO^\prime}^2
      \leftidx{^m}S
      \leftidx{^n}S
      {\BO}^2\leftidx{^m}\GO 
      \leftidx{^n}\GO
  \end{eqnarray}\vspace{-0.5cm}
  \begin{eqnarray*}\label{eq:F12b}
  \nonumber
    \left\{\leftidx{^n}f ^2\left[ \frac{x-\!\leftidx{^n}x}{\mu_0} 
    \vect{\hat{x}}                 
    +\frac{y-\!\leftidx{^n}y}{\mu_0} 
    \vect{\hat{y}}\right]
+    \leftidx{^m}f ^2\left[ \frac{x-\!\leftidx{^m}x}{\mu_0} 
    \vect{\hat{x}}                 
    +\frac{y-\!\leftidx{^m}y}{\mu_0} 
    \vect{\hat{y}}\right]\right\}.
  \end{eqnarray*}
  \begin{figure*}
  \centering
  \includegraphics[trim=3.0cm 0.0cm 2.0cm 1.2cm, clip=true, width=0.95\linewidth]{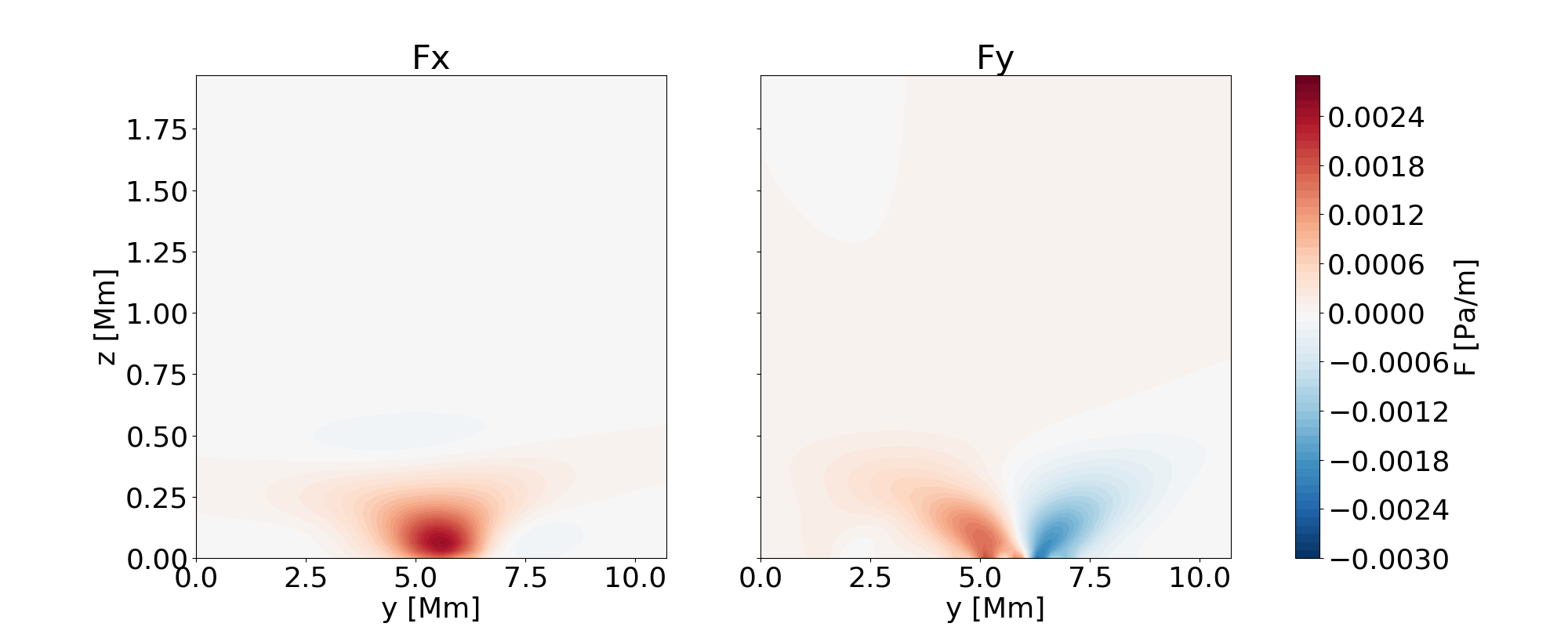}
  \caption{
  2D slice \rev{of model horizontal balancing forces} at $x=8.15$ Mm \rev{as defined by Equation\,\eqref{eq:Fbalsum}}.
  \label{fig:Fterms}}
  \end{figure*}
  If we generalise to a system of $N$ flux tubes with $\vect{B}_b =
  \leftidx{^1}{\vect{B}_b}+\leftidx{^2}{\vect{B}_b}+...
  +\leftidx{^N}{\vect{B}_b}$, then the pressure can be fully described by 
  \begin{equation}\label{eq:pmn}
  p_b=p_b\vv+ \sum_{m=1}^N\leftidx{^m}p_b\hh+
              \sum_{m,n=1|_{n>m}}^N\leftidx{^{mn}}p_b\hh,
  \end{equation}
  in which $p_b\vv$ is derived from the interpolated observed profile, constrained to be monotonically decreasing with height, and $\leftidx{^m}p_b\hh$ is defined by Equation\,\eqref{eq:px}.
  The pressure adjustment due to each pairwise flux tube interaction is given by 
  \begin{equation}\label{eq:apx}
    \leftidx{^{mn}}{p_b\hh} =
      \frac{
      \leftidx{^m}S
      \leftidx{^n}S
      \fO
      }{2\mu_0}
      \leftidx{^m}\GO 
      \leftidx{^n}\GO 
      \left[
      {\BO^\prime}^2
      +\BO
      \BO^{\prime\prime}
      \right]
    -\frac{\leftidx{^m}{{B}_{bz}}\leftidx{^n}{{B}_{bz}}}{\mu_0}.
  \end{equation}
  The corresponding expression for the plasma density is   
  \begin{equation}\label{eq:rhosum}
  \rho_b=\rho_b\vv+ \sum_{m=1}^N\leftidx{^m}\rho_b\hh+
              \sum_{m,n=1|_{n>m}}^N\leftidx{^{mn}}\rho_b\hh,
  \end{equation}
  in which $\rho_b\vv$ is the product of $g^{-1}$ and the $z$-derivative of $p_b\vv$, and $\leftidx{^m}\rho_b\hh$ is defined by Equation\,\eqref{eq:rho1}.
  The density adjustments due to each pairwise flux tube interaction are given
  by
  \begin{eqnarray}\label{eq:rhomn}
    \leftidx{^{mn}}{\rho_b\hh}\,
    =&&
    \hspace{-0.5cm}2\frac{\leftidx{^{m}}{{S}}\leftidx{^{n}}{{S}}}{\mu_0\,g}
      \leftidx{^m}\GO 
      \leftidx{^n}\GO 
      \BO\BO^{\prime}
    \left[
    \left(
    \frac{\leftidx{^{m}}{f} +\leftidx{^{n}}{f}}{\fO} -2 
    \right)
      {\BO}^2
    \right.
    \\\nonumber
    -&&
       \hspace{-0.5cm}\frac{ \leftidx{^m}f+\leftidx{^n}f}{2}
    \left(
      \frac{{\BO^\prime}^2}{{\BO}^2} + \frac{\BO^{\prime\prime}}{\BO}
    \right)
    +
      \frac{
      \fO
      }{4}
      \left(
      3\frac{\BO^{\prime\prime}}{\BO}
      +
      \frac{\BO^{\prime\prime\prime}}{\BO^\prime}
      \right)
    \\\nonumber
        +&&\hspace{-0.5cm}\left\{
    {(x-\leftidx{^m}x )(x-\leftidx{^n}x )}
    +
    {(y-\leftidx{^m}y )(y-\leftidx{^n}y )}{}
    \right\}
    \\\nonumber
    &&
      \hspace{-0.5cm}\left.
      \left\{
      \left(
     1-\frac{\leftidx{^{m}}{f}+\leftidx{^{n}}{f}}{\fO} 
      \right) 
    {\BO^\prime}^2 + \BO\BO^{\prime\prime}
    -2
      \frac{{\BO}^4}{\fO}
    \right\}
    \right].
  \end{eqnarray}
  The net balancing force in Equation\,\eqref{eq:delP} is then fully specified as
  \begin{equation}\label{eq:Fbalsum}
  {\vect{F}}_{\rm bal}=
              \sum_{m,n=1|_{n>m}}^N\leftidx{^{mn}}{\vect{F}}_{\rm bal}.
  \end{equation}
  
  \begin{figure*}
  \centering
  \includegraphics[trim=1.2cm 0.2cm 0.0cm 2.85cm, clip=true, height=0.68\linewidth, width=0.72\linewidth]{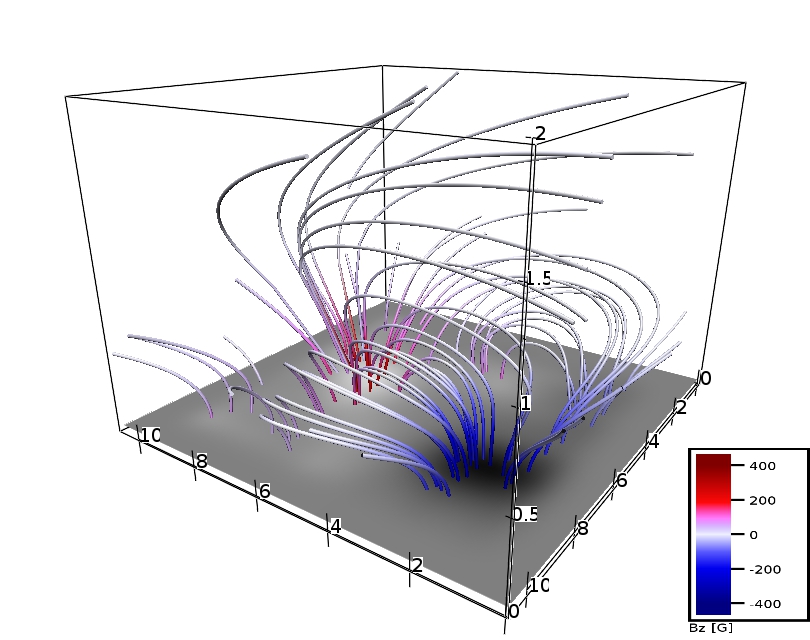}
  \caption{
  3D plot of chromospheric loop reconstruction. \rev{Colour indicates the magnetic fieldline vertical component and grey scale $B_z$ at the photosphere.}
  \label{fig:3dhmi}}
  \end{figure*}
  \begin{figure*}
  \centering
  \includegraphics[trim=2.5cm 0.4cm 1.3cm 1.4cm, clip=true, width=0.95\linewidth]{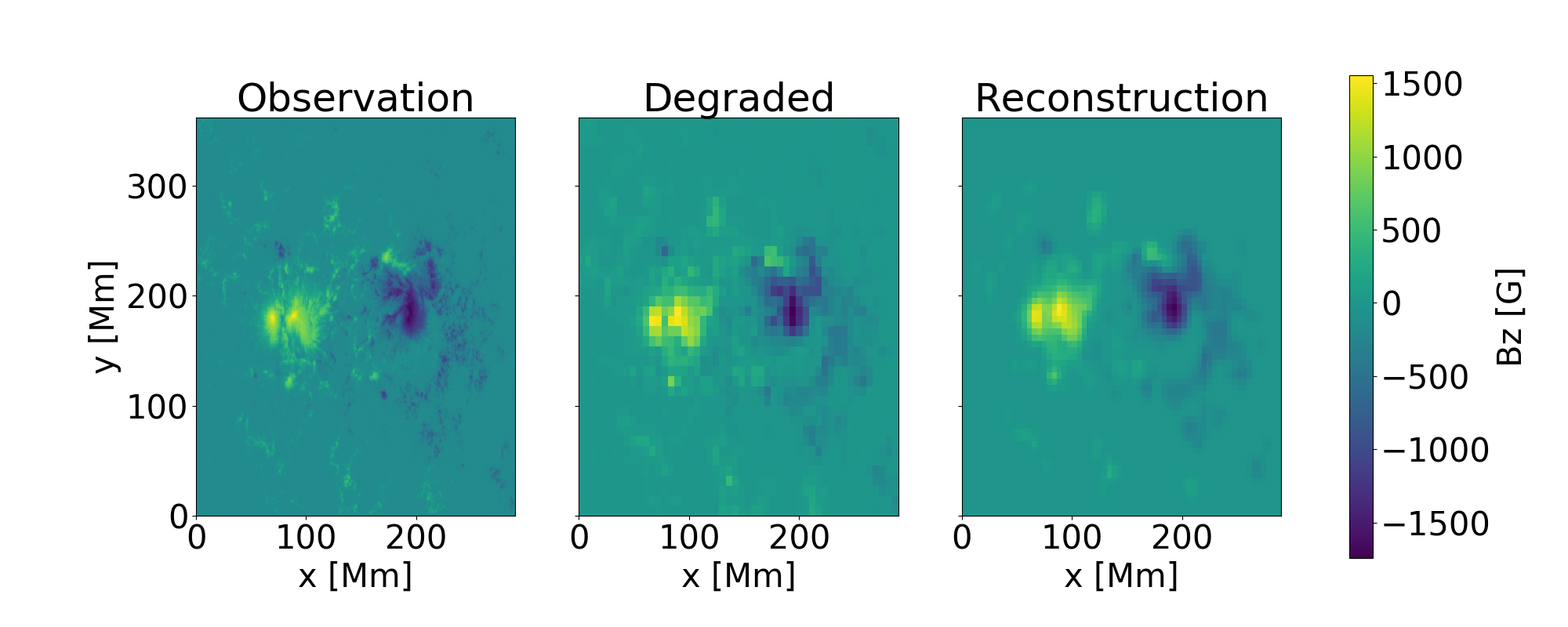}
  \caption{
  HMI magnetogram (left), spatially degraded HMI (centre) and model photosphere (right).
  \label{fig:hmimapc}}
  \end{figure*}
 
   \begin{figure*}
  \centering
  \includegraphics[trim=5.4cm 0.15cm 0.05cm 4.8cm, clip=true, height=0.65\linewidth, width=0.65\linewidth]{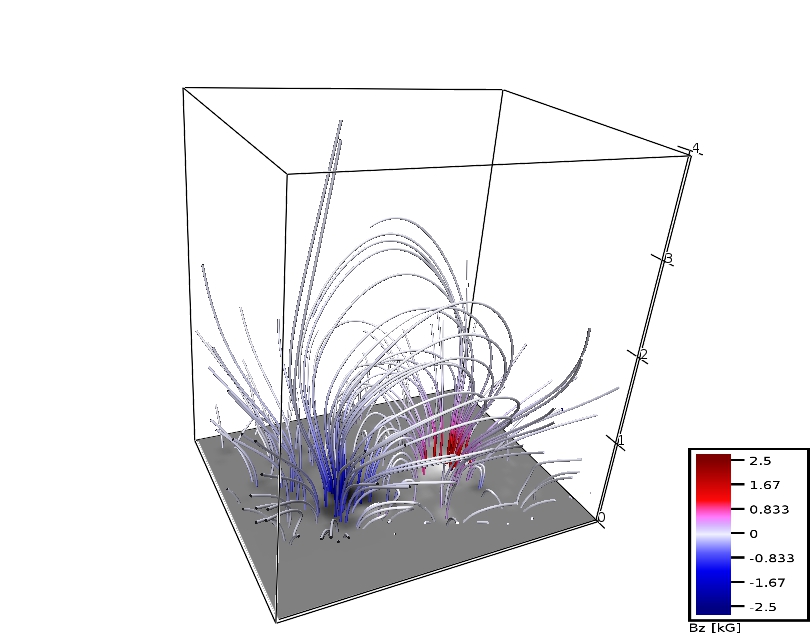}
  \caption{
  3D plot of magnetic fieldlines above an active region, including the lower corona.
  \rev{Colour indicates their vertical component. Grey scale shading shows $B_z$ at the photosphere. Units are in kG, not G as in Figure\,\ref{fig:3dhmi}.}
  \label{fig:3dhmic}}
  \end{figure*}

  \section{Application of the Model}\label{sect:apps}

  \subsection{Fitting arbitrary flux tubes\label{sect:arb}}
  A stable atmosphere can be generated for any distribution of photospheric magnetic field by using the observed magnetic field in each pixel to construct a series of interacting flux tubes. To demonstrate this, the atmosphere is constructed using a subsection of the HMI magnetogram observed on 2014.07.06\_00\_00\_45. A relatively small region ($16\times16$ pixels) is chosen that features a few isolated magnetic regions of opposite polarity.
  
  In a numerical grid of horizontal dimension $64\times64$ magnetic flux tubes with $f_0\simeq750$\,km are fitted for each pixel in the observing box. Figure\,\ref{fig:hmimap} shows the observed HMI magnetogram (left) and the reconstructed photospheric magnetic field (right). A region around the observation is set to zero to allow numerical boundaries to be well defined when the atmosphere is used for simulations. 
  {Shown in} Figure\,\ref{fig:hmimap}, there is strong  agreement between the observation and the reconstruction both in terms of locations and magnitude of magnetic field.

  The density and pressure modifications{,} required to stabilise the magnetic field{, are} generated using the methods outlined in Section\,\ref{sect:loop}.
  The additional forcing terms $\vect{F}_{\rm bal}$ applied to account for the magnetic tension effects between neighbouring flux tubes are plotted in Figure\,\ref{fig:Fterms}. 
  The forcing terms are significant only in the lower atmosphere and are zero in most of the domain. For context, the magnitude of the forcing terms is maximally around 2\% of the horizontal pressure gradient. 
  {T}hese forcing terms represent a small adjustment to the system.
  
  The end result is a 3D FME that models the photospheric magnetic field, shown in Figure\,\ref{fig:3dhmi}, using VAPOR \citep{clyne05,clyne07}.
  Due to the modest footpoint magnetic field of around 30\,mT,
  the loop is mainly confined to the chromosphere, so we model the region to a height of 2\,Mm above the photosphere.
  Simulations of a {well-observed region, in preparation, aim to illustrate the model's effectiveness for such complex networks}.


Above active regions, the magnetic field can easily extend through the transition region and into the solar corona. To test the construction of such atmospheres we apply the same methodology to an active region with vertical magnetic field strength of $B_z \approx \pm 2500$ G. This region is much larger than the previous test and hence fitting a flux tube to each observational pixel is computationally expensive. To circumvent this, we degrade the observation to a lower spatial resolution (see Figure\,\ref{fig:hmimapc} and fit flux tubes to the strongest sources only, yielding the network plotted in Figure\,\ref{fig:3dhmic}.

\section{Results summary}

In this article we describe and demonstrate a new method for reconstructing a st{ationary} state solar atmosphere, with realistic magnetic configuration. 
The model parameters have been streamlined and generalised, making them easy to apply for arbitrary photospheric magnetic field sources.
Calculating the magnetic fields and resulting atmosphere is computationally efficient, available in parallel python from PYSAC {(}{\href{https://github.com/fredgent/pysac}{https://github.com/fredgent/pysac}}{)}.

The free parameters in radial scaling and scale height, and the generalised inclusion of any ambient atmosphere models, makes the method versatile for a number of scientific problems.
The physical veracity of the parameters can, however, be constrained by comparison with observations of the magnetic field and kinetics at various heights. {The stability of the solution can also be confirmed by numerical simulation for each configuration. This was carried out for the flux-tube pair solution used in \citet{SFGVE18}, by treating the solution as the MHD perturbations, and the system remained stationary to within machine accuracy.}

{We provide a new method to extrapolate the magnetic field from observations {in the lower solar atmosphere}. A common approach to
obtaining a steady state magnetic configuration is to start with a potential field extrapolation of vertical magnetic field measurements {\citep{SdR03}}. This is then evolved in MHD simulations to find an equilibrium \citep[{e.g.,}][]{GN05,HHPC10,FSE11,HGPC15}. The new construction method does not depend on any {Dirichlet nor von Neumann type boundary conditions or timestep constraints required for the MHD PDE solver}. It does require care in the choice of parameters to avoid unrealistic gas density or temperatures. It may be a faster technique, but we propose to compare these two methods in future work. Firstly, the efficiency of deriving the steady state atmosphere using both methods shall be measured. Secondly, the results of simulations using each construction of a solar region shall be compared with the photospheric and chromospheric observations.
}
%
  \section*{Acknowledgements}
  The authors wish to acknowledge CSC – IT Center for Science, Finland, for computational resources and the financial support by the Academy of Finland to the ReSoLVE Centre of Excellence (project no. 307411). {FAG, RE and VF were supported by STFC Grant R/131168-11-1.}
  BS is supported by STFC research grant ST/R000891/1.
  We would also like to thank HPC-EUROPA3 Transnational Access Program for providing HPC facilities and support.
  \rev{We thank the anonymous referee for constructive criticism and advice.}

  \bibliographystyle{mnras}      
  \bibliography{refs}


  \label{lastpage}
 
\end{document}